\documentclass[12pt]{article}
\usepackage{fontenc}
\usepackage[latin1]{inputenc}
\usepackage[english]{babel}
\usepackage{graphicx,amssymb,MnSymbol,amsmath,bm,lineno,color}
\usepackage[normalem]{ulem}

\usepackage{xfrac,setspace}

\usepackage{mathtools}

\begin{document}

\title{The geometrization of electromagnetism}

\author{Celso de Araujo Duarte\\
Departamento de F\'isica, Universidade Federal do Paran\'a\\
CP 19044 -- 81531-990, Curitiba (PR), Brazil
\\E-mail: celso.duarte@ufpr.br}

\maketitle

\hyphenation{re-la-ti-vis-tic}

\abstract{Previous work \cite{D1,D2,D3} expose a synthetic, Riemannian geometrization of electromagnetism that emerges naturally from the experimental bases provided by the Lorentz force and the Maxwell equations.

Since the components of the associated space-time metric tensor depend on the four-vector velocity, Riemannian geometry does not fully describe the space-time geometry.

The aim of the present work is to provide the suitable geometrical basis for the problem, applying a \textit{pseudo-Finsler geometry} -- a Finsler geometry with non homogeneous norm.}


\section{Introduction}\label{I}

Michael Faraday was strongly convinced that electromagnetism and gravitation were linked, yet performed an unsuccessful experimental test \cite{Far}. Despite fruitless, it did not extinguish his faith in his own conjecture: ``(...)\textit{ do not shake my strong feeling of the existence of a relation between gravity and electricity, though they give no proof that such a relation exists}''.

The advent of general relativity theory (GRT) motivated the search for a geometric origin of electromagnetism that could eventually lead to the desired unification.

Few years after the GRT was presented, we had the emergence of the scale invariant Weyl gauge theory in 1918 \cite{Weyl}; the five-dimension Kaluza \cite{Kal} and Klein \cite{Kle} theories, between 1921-1926; the affine field theory of Schr\"odinger \cite{Schr,Hla}; the asymmetric metric works of Eddington (1921) \cite{Edi}, Einstein \cite{Ein}, Schr\"odinger, Strauss and Kaufman (along the 1940s - 1950s); the asymmetric model of Randers (1941) \cite{Ran}; also, the Einstein-Cartan theory that incorporates the space-time torsion (1920s) \cite{Car} reloaded by Sciama and Kibble in the 1960s, amongst others \cite{Goen}.

The Kaluza-Klein multidimensional theory dominated, serving as the basis for the construction of modern quantum theories such as superstring theory. The current mainstream lies in the construction of a quantum theory of gravitation, and classical theories have been mostly abandoned.

In any case, the most successful and realistic mathematical model certainly stands in the largest set of established experimental results that, in the realm of electromagnetism, are the Lorentz force and Maxwell equations.

In this sense, the present work continues the recent proposal of a classical geometrization of electromagnetism, initially presented in reference \cite{D1} (2014) and continued in references \cite{D2} and \cite{D3}. The initial step started from a reinterpretation of the terms of the geodesic equation for a massive charged test particle under the action of an external electromagnetic field. This requirement led to the following space-time metric tensor\footnote{Reserving the letter $g$ for the gravitational contribution to the metric tensor, we preferred using the letter $h$ to represent the electromagnetic contribution.} (reference \cite{D1}, expression 15 therein):
\begin{equation}\label{hmn0}
\left\{h_{\mu\nu}\right\}=\left(
\begin{array}{cc}
1+2r\phi_0 & -r\boldsymbol{\varphi}\\
-r\boldsymbol{\varphi} & -\mathbf I
\end{array}
\right),
\end{equation}
where $\left\{\varphi_0,\boldsymbol{\varphi}\right\}=\left\{\varphi_{\mu}\right\}$ are the covariant components of the four-vector electromagnetic potential, and $r=q/m$ is the charge ($q$) to mass ($m$) ratio of the test particle\footnote{In synthetic notation, the right member of \ref{hmn0} is a 4$\times$4 matrix, being $\boldsymbol{\varphi}$ a three-component line vector at the first line, and a three component column vector at the first column. $\mathbf I$ is the 3$\times$3 identity matrix.}. This metric, when substituted in the geodesic equation, agrees with the expected Lorentz force.

Given the metric \ref{hmn0} and the Maxwell equations, the following field equation was constructed with the Ricci curvature $R_{\mu\nu}$ and the Ricci scalar $R$ :
\begin{equation}\label{Field}
R_{\mu\nu}-\frac12g_{\mu\nu}R=\lambda Q_{\mu\nu},
\end{equation}
where $Q_{\mu\nu}$ is a charge-current-dependent magnitude, and $\lambda=\mu_0q/m$. Equation \ref{Field} is formally equivalent to the Einstein field equation of gravitation (EFE).

Actually, expressions \ref{hmn0} and \ref{Field} are not of a tensor nature. Meanwhile this issue was not solved, a second work \cite{D2} introduced a new strategy, with the following algebraic manipulation with the infinitesimal space-time interval $ds^2=h_{\mu\nu}dx^{\mu}\nu^{\nu}$ (using expression \ref{hmn0}),
\begin{equation}\label{ds}
ds^2=h_{\mu\nu}dx^{\mu}dx^{\nu}=\eta_{\mu\nu}dx^{\mu}\nu^{\nu}+2r\varphi_{\nu}dx^{\nu}ds,
\end{equation}
where it was used the identity $ds=u_{\mu}dx^{\mu}$. It was provided an interpretation for expression \ref{ds} in the context of quantum mechanics encompassing the concept of quantum action, showing that \textit{the main effect of the electromagnetic field is to change the length of the space-time interval, while} (as yet demonstrated in reference \cite{D1}) \textit{the Riemannian scalar curvature $R$ is zero}.

From \ref{ds}, the following approximation was obtained, up to first order in the components of $\varphi_{\mu}$:
\begin{equation}\label{ds1}
ds=\sqrt{\eta_{\mu\nu}dx^{\mu}dx^{\nu}}+r\varphi_{\nu}dx^{\nu}.
\end{equation}
In the above expression, the replacement of the Minkowski metric $\eta_{\mu\nu}$ with the gravitational one, $\eta_{\mu\nu}\rightarrow g_{\mu\nu}$, gives the metric proposed by Gunnar Randers \cite{Ran}.


Returning to the problem of the lack of tensor character in reference \cite{D1}, it was solved in reference \cite{D2} with the following full-tensor metric:
\begin{equation}\label{hmn}
h_{\mu\nu}(x,u)=\eta_{\mu\nu}+r\left(u_{\mu}\varphi_{\nu}+u_{\nu}\varphi_{\mu}\right).
\end{equation}
Correspondingly, the Ricci curvature $R_{\mu\nu}$ and Ricci scalar $R$, and the expression \ref{Field}, were presented with tensor form.

Yet the metric \ref{hmn} is not Riemannian, owing to the dependence on the velocity $u^{\alpha}$. As in the case of Randers metric, the well-suited geometry for such a situation is the Finsler geometry (FG). The explicit dependence of the metric \ref{hmn} with respect to the four-velocity $u^{\mu}$ implies that the length of paths depends on the velocity and on the direction, bringing a new panorama that is not fitted by the framework of the RG. The use of RG in such a physical situation results in a loss of information associated with directionality dependence. In this situation, we enter the domain of FG \cite{Fin,Pfei,Lam,S1,S2,SS}, which is the best suited.

The object of the present work is to apply the FG to electromagnetism. If by one side, the RG-based references \cite{D1} and \cite{D3} brougth us the field equations of the type:
\begin{equation}\label{fe}
\begin{array}{ccc}
\underbrace{R_{\mu\nu}-\dfrac12h_{\mu\nu}R} & \propto &\underbrace{ Q_{\mu\nu}},\\
\colorbox{yellow}{\textsc{geometry}} & &
\makebox[25pt][c]{
\colorbox{green}{$\begin{array}{c}
\textsc{charge}\\
\textsc{source}
\end{array}$}
}\end{array}
\end{equation}
where $Q_{\mu\nu}$ contains charge and current densities, the corrections brought by the formalism of the present work provide:
\begin{equation}\label{fef}
\begin{array}{ccccc}
\colorbox{yellow}{\textsc{geometry}} & \propto &
\makebox[45pt][c]{
\colorbox{green}{
$\begin{array}{c}
\textsc{charge}\\
\textsc{source}
\end{array}$}}
&+&
\makebox[33pt][c]{
\colorbox{magenta}{
$\begin{array}{c}
\textsc{forces}
\end{array}$}}
\hspace{8pt}+\hspace{11pt}
\makebox[55pt][c]{
\colorbox{cyan}{
$\begin{array}{c}
\textsc{stress-energy}\\
\textsc{tensor}
\end{array}$}}
\end{array}
\end{equation}
The detailed explanation of the meaning of the terms named in \ref{fef} will be provided in the present work.

Foundations of FG are presented in Section \ref{stFG}, which is then applied to the geometrization of electromagnetism in Section \ref{Fge}. Section \ref{Compar} is dedication to discussions and analyses, and the Conclusions are at the end of the article. A final Appendix is dedicated to presenting some significant expressions in the context.

In summary, it was constructed the geometrization of the electromagnetism, where we have a relative, test-particle-dependent geometry, strongly based in supporting experimental basis, successively presented and refined since the initial reference \cite{D1}. This geometrization is a good candidate for the geometric description of electromagnetism.


\section{Space-time curvature in Finsler geometry}\label{stFG}

In FG, a norm is defined rather than the scalar product of RG. This norm is a non-negative function $\mathcal F(x,y)$ of the set of coordinates $\left\{x^{\mu}\right\}$, here represented in compact form as $x$, that acts on a vector $y$ taken in the tangent bundle, here represented by $\dot{x}=\left\{\dot{x}^{\mu}\right\}$.

Then, given the vector $\dot x \neq 0$ in a Finsler manifold, the Finsler metric $g_{\mu\nu}$ is given by the Hessian matrix of $\mathcal F^2$,
\begin{equation}\label{metric}
g_{\mu\nu}(x,\dot x)=\frac 12\frac{\partial^2 \mathcal F^2(x,\dot x)}{\partial \dot x^{\mu}\partial \dot x^{\nu}},
\end{equation}
where we see the explicit dependence of $g_{\mu\nu}$ with respect to both the coordinates $x$ and the tangent vector $\dot x$. In contrast, in RG, the metric depends only on the coordinates. For this reason, FG differs from RG in the sense that lengths depend not only on coordinates (Riemannian case) but also on directions.

The counterpart of expression \ref{metric} is
\begin{equation}\label{F}
\mathcal F^2(x,\dot x)=g_{\mu\nu}(x,\dot x) \dot x^{\mu}\dot x^{\nu}.
\end{equation}
In some references, the factor $\frac12$ in expression \ref{metric} is moved to \ref{F}, in order that the second member of \ref{F} results in a formal similarity to a kinetic energy.

The line element $ds$ is given by
\begin{equation}\label{ds2ds}
ds^2=g_{\mu\nu}(x,\dot x) dx^{\mu}dx^{\nu},\ \ \ \ \ \ \ ds=\sqrt{g_{\mu\nu}(x,\dot x) dx^{\mu}dx^{\nu}},
\end{equation}
and $s$ is the independent variable in which is based the set of all parametric curves $\left\{x^{\mu}(s),\dot x^{\mu}(s)\right\}$.

Now, from \ref{F},
\begin{equation}\label{Fds}
\mathcal F(x,\dot x)=\sqrt{g_{\mu\nu}(x,\dot x) \dot x^{\mu}\dot x^{\nu}}.
\end{equation}
The length is then defined in terms of the functional
\begin{equation}\label{L}
\mathcal L=\int^b_a \mathcal F\left(x(s),\dot x(s)\right) ds,
\end{equation}
which is parametrized by $s$, and the trajectory from a point $a$ to a point $b$ is defined by a parametric curve $\left\{x(s)\right\}$. The directional dependence is determined by the dependence of the norm $\mathcal F(x,\dot x)$ with respect to the vector field $\dot x$ that is absent in the Riemannian metric. A consequence of this is the fact that, while in the RG the value of the integral \ref{L} is the same for both directions, $a\rightarrow b$ or $b\rightarrow a$, the same does not apply to the Finslerian case.

While in the RG, the Riemann and the Ricci curvature tensors are functions of the Christoffel symbols \cite{Lan1,Whe}, in FG we define the Ricci curvature as a function of the \textit{spray} coefficients $G^{\mu}$ that are given by \cite{S1,S2,SS,VN}:
\begin{equation}\label{Gi}
G^{\mu}(x,\dot x)=\frac 14 g^{\mu\alpha}(x,\dot x)\left(\frac{\partial^2 \mathcal F^2(x,\dot x)}{\partial x^{\beta}\partial \dot x^{\alpha}}\dot x^{\beta}-\frac{\partial \mathcal F^2(x,\dot x)}{\partial x^{\alpha}}\right).
\end{equation}
The Ricci curvature tensor in FG, $\mathcal R_{\mu\nu}$, is given as a function of $G^{\mu}$ by (calligraphic $\mathcal R$ represents the curvatures in FG, and Roman print in RG)
\begin{equation}\label{Rmn}
\begin{array}{ll}
\mathcal R^{\mu}{}_{\nu}(x,\dot x)=& 2\dfrac{\partial G^{\mu}(x,\dot x)}{\partial x^{\nu}}-\dot x^{\alpha}\dfrac{\partial^2 G^\mu(x,\dot x)}{\partial x^{\alpha}\partial \dot x^{\nu}}+2G^{\alpha}(x,\dot x)\dfrac{\partial^2 G^{\mu}(x,\dot x)}{\partial \dot x^{\alpha}\partial \dot x^{\nu}}-\\
\\
&\dfrac{\partial G^{\mu}(x,\dot x)}{\partial \dot x^{\alpha}}\dfrac{\partial G^{\alpha}(x,\dot x)}{\partial \dot x^{\nu}}.
\end{array}
\end{equation}
From this, we obtain the Ricci curvature $\mathcal R$ which is a scalar,
\begin{equation}\label{Ric}
\mathcal R(x,\dot x)=\mathcal R^{\alpha}{}_{\alpha}.
\end{equation}
Minimization of the functional $L$ provides the geodesic equation given by
\begin{equation}\label{geod}
\frac{d^2 x^{\mu}}{d s^2} + 2G^{\mu}(x,\dot x)=0.
\end{equation}
In particular, the curvature provided by expression \ref{Rmn} reduces to
\begin{equation}\label{RmnF}
\mathcal R^{\mu}{}_{\nu}=R^{\mu}{}_{\alpha\nu\beta}u^{\alpha}u^{\beta},
\end{equation}
where $R^{\mu}{}_{\alpha\nu\beta}$ is the well-known Riemann tensor of RG. 

The RG can be considered a particular case of the Finslerian one. However, note that the Ricci curvature presented in expression \ref{Rmn} is not the same as the Ricci curvature on the context of RG and general relativity: in FG, the Ricci tensor is obtained by contraction of the second and fourth indexes of the Riemann tensor $R^{\mu}{}_{\alpha\nu\beta}$ with the four-velocity, as shown in \ref{RmnF}, while in RG, the Ricci tensor is obtaned by self contraction of these same indexes of the Riemann tensor, i.e. $R^{\mu}{}_{\nu}=g^{\alpha\beta}R^{\mu}{}_{\alpha\nu\beta}=R^{\mu}{}_{\alpha\nu}{}^{\alpha}$. The latter is the Ricci tensor that appears in the EFE of gravitation.

An interesting fact is the link between the Finslerian Ricci tensor \ref{Rmn}, $\mathcal R^{\mu}{}_{\nu}$, and the Riemannian one, $R^{\mu}{}_{\nu}$: contracting the free indexes of \ref{RmnF}, we obtain the Ricci scalar \ref{Ric}, and
\begin{equation}\label{Ric2}
\mathcal R(x,\dot x)=\mathcal R^{\sigma}{}_{\sigma}=R^{\sigma}{}_{\alpha\sigma\beta}u^{\alpha}u^{\beta}=R{}_{\alpha\beta}u^{\alpha}u^{\beta}.
\end{equation}


\section{The geometrization of electromagnetism}\label{Fge}

\subsection{Preliminary}\label{prelim}

In the present section, we apply the formalism of the FG presented in the previous section to our model represented by the metric \ref{hmn}.

Starting with this metric \ref{hmn}, we replace $g_{\mu\nu}$ with $h_{\mu\nu}$ in \ref{ds2ds}, and write the infinitesimal space-time interval:
\begin{equation}\label{ds2}
ds^2=\left(\eta_{\mu\nu}+2ru_{\mu}\varphi_{\nu}\right)dx^{\mu}dx^{\nu}.
\end{equation}
Then, we obtain directly the norm from \ref{Fds}, using the identity $u_{\mu}u^{\mu}=1$:
\begin{equation}\label{Fh2}
\mathcal F(x,u)=\sqrt{\eta_{\mu\nu}u^{\mu}u^{\nu}+2r\varphi_{\nu}u^{\nu}},
\end{equation}
Note that, for this norm we have the corresponding Lagrangian
\begin{equation}\label{Lagr}
L=\frac12 m\mathcal F^2,
\end{equation}
and the associated canonical momentum is
\begin{equation}\label{mom}
P_{\alpha}=\frac{\partial L}{\partial u^{\alpha}}=m\left(u_{\alpha}-r\varphi_{\alpha}\right)=p_{\alpha}-q\varphi_{\alpha},
\end{equation}
which is in accordance with the minimal coupling of the electromagnetic field.

This norm is not a homogeneous function of the velocity since, for a given $\lambda\in \mathbb{R}$, there is no a $\theta\in\mathbb R \mid \mathcal F(x,\lambda u)=\lambda^{\theta}\mathcal F(x,u)$ is satisfied. For this reason, we call this a ``pseudo-Finsler geometry'' (PFG). e may address this problem thinking about the two limit cases: when the term $\eta_{\mu\nu}u^{\mu}u^{\nu}$ in \ref{Fh2} is dominant over the term $2r\varphi_{\nu}u^{\nu}$, $\mathcal F^2$ scales approximately quadratically with the velocity $u$ (regime of low fields and/or low velocities); in the opposite case, $\mathcal F^2$ scales approximately linearly with the velocity $u$ (high fields and/or velocities). However, when determining the spray coefficients $G^{\mu}$, the Minkowskian part $\eta_{\mu\nu}u^{\mu}u^{\nu}$ will drop out through the derivation with respect to the coordinates (the derivatives $\partial_{\beta}$ and $\partial_{\alpha}$ in expression \ref{Gi}), and for this reason both limit cases can be treated equally under the mathematical aspect\footnote{In particular, it is worth mentioning that the original metric presented by Randers in 1941 corresponded to the following space-time interval $ds$ \cite{Ran}:
$$ds=k_{\mu}dx^{\mu}+\sqrt{g_{\mu\nu}dx^{\mu}dx_{\nu}}.$$
Dividing by $ds$, the norm is:
$$\mathcal F=\frac{ds}{ds}=k_{\mu}u^{\mu}+\sqrt{g_{\mu\nu}u^{\mu}u_{\nu}}\equiv 1.$$
The norm \ref{Fh2} is similar to the Randers one, but it is not homogeneous, while the latter is homogeneous (of degree 1).}

We may get around the inhomogeneity in the following way: take \ref{ds2} and add $(\varphi_{\nu}dx^{\nu})^2$ to both members, group terms in a squared binomial, and obtain the result yet shown in reference \cite{D2} (equation 29 therein):
\begin{equation}\label{ds3}
ds=\sqrt{\eta_{\mu\nu}dx^{\mu}dx^{\nu}+r^2\left(\varphi_{\mu}dx^{\mu}\right)^2}+r\varphi_{\nu}dx^{\nu},
\end{equation}
consequently, the norm can be written as:
\begin{equation}\label{Fh3}
\mathcal F=\sqrt{\eta_{\mu\nu}u^{\mu}u^{\nu}+(r\varphi_{\nu}u^{\nu})^2}+r\varphi_{\nu}u^{\nu}\ \ \longrightarrow \ \ \mathcal F(x,u)=\alpha(x,u)+\beta(x,u)
\end{equation}
where
\begin{equation}\label{alphbet}
\alpha=\sqrt{\eta_{\mu\nu}u^{\mu}u^{\nu}+\beta^2}, \ \ \ \ \ \beta=r\varphi_{\nu}u^{\nu},
\end{equation}
that is the standard presentation of Randers metrics. This last result is homogeneous of degree 1 in the velocities.

We will study separately both expressions \ref{Fh2} and \ref{Fh3} as two individual cases: in the Case I, next subsection \ref{CI}, we will study the situation derived from expression \ref{Fh2}; and in the Case II, at subsection \ref{CII}, we will study the situation derived from expression \ref{Fh3}. The comparative analysis will be presented in Section \ref{Compar}.


\subsection{Case I: $\mathcal F=\sqrt{\eta_{\mu\nu}u^{\mu}u^{\nu}+2r\varphi_{\nu}u^{\nu}}$}\label{CI}

For the norm given by \ref{Fh2}, evaluation of the spray coefficients $G^{\mu}$ is simple, and gives exactly:
\begin{equation}\label{Gih}
G^{\left\{I\right\}\mu}(x,u)=\frac12ru^{\beta}F_{\beta}{}^{\mu}, 
\end{equation}
where $F_{\beta\alpha}=\partial_{\beta}\varphi_{\alpha}-\partial_{\alpha}\varphi_{\beta}$ is the electromagnetic tensor (the upper index $\left\{I\right\}$ labels the present Case I). Note that a canonical Finsler spray is homogeneous of degree two, while \ref{Gih} is of degree one. Apropos, the inverse metric tensor $h_{\mu\nu}$ is
\begin{equation}\label{hmninv}
h^{\mu\nu}(x,u)=\eta_{\mu\nu}+r\left(u_{\mu}\varphi_{\nu}+u_{\nu}\varphi_{\mu}\right)+\mathcal{O}^2\left(u^2,\varphi^2\right).
\end{equation}

\subsubsection{The geodesic equation, Case I}

Placing $G^{\mu}$ in the geodesic equation,
\begin{equation}\label{geod1}
\frac{du^{\mu}}{ds}+2G^{\mu}=0
\end{equation}
gives:
\begin{equation}\label{geod2}
\frac{du^{\mu}}{ds}=ru^{\beta}F_{\mu}{}^{\beta},
\end{equation}
and this is exactly the equation of movement for a test charge \cite{D1,D2,D3}. Note that a degree two spray would provide a squared velocity-dependence that is not in accordance with the experimental well-established result. 

\subsubsection{Curvatures, Case I}

We may determine the curvature, substituting \ref{Gih} in expression \ref{Rmn}. Since $G^{\mu}$ is linear in $u$, the second derivative term (third term in the second member of \ref{Rmn}) is zero. Using the Bianchi identity for the electromagnetic tensor, $\partial_{[\alpha}F_{\beta\gamma]}=0$, we arrive to:
\begin{equation}\label{Rmn3}
\mathcal R^{\left\{I\right\}}_{\mu\nu}= ru^{\alpha}F_{\alpha(\mu,\nu)}+\frac14r^2F_{\mu\alpha}F_{\nu}{}^{\alpha},
\end{equation}
where $F_{\alpha(\mu,\nu)}=\frac12\left(F_{\alpha\mu,\nu}+F_{\alpha\nu,\mu}\right)$.

Now we determine the Ricci scalar, $\mathcal R(x,u)$, from the trace of $R^{\mu}{}_{\nu}(x,u)$, and using the Gauss and the Amp\`ere laws, we arrive to:
\begin{equation}\label{Rs}
\mathcal R^{\left\{I\right\}}=-r\mu_0 u^{\alpha}j_{\alpha}+\frac14r^2 F_{\alpha\beta}F^{\alpha\beta}.
\end{equation}

\subsubsection{The field equation, Case I}

Now we construct the field equation of electromagnetism, based in expressions \ref{Rmn3} and \ref{Rs}. The noticeable contracted double product of components of the electromagnetic tensor in these equations are just the parts of the electromagnetic energy-stress tensor $T_{\mu\nu}=\mu_0^{-1}(F_{\mu\alpha}F_{\nu}{}^{\alpha}-\frac14h_{\mu\nu}F_{\alpha\beta}F^{\alpha\beta})$, with which we construct the \textit{ad hoc} the tensor $\mathcal{E}_{\mu\nu}$ and the geometric electromagnetic field equation: 
\begin{equation}\label{E}
\begin{array}{lllll}
\mathcal{E}^{\left\{I\right\}}_{\mu\nu}&=&\mathcal R^{(I)}_{\mu\nu}&-&\dfrac14h_{\mu\nu}\mathcal R^{(I)}=\\
\\
&&&&ru^{\alpha}\left(\dfrac14h_{\mu\nu}\mu_0j_{\alpha}+F_{\alpha(\mu,\nu)}\right)+\dfrac14r^2\mu_0T_{\mu\nu}
\end{array}
\end{equation}
where $\mathcal{E}_{\mu\nu}^{(I)}$ is the case I electromagnetic equivalent to the gravitational Einstein tensor $G_{\mu\nu}=R_{\mu\nu}-\frac12g_{\mu\nu}R$. The last expression can be simplified, considering that $u^{\alpha}j_{\alpha}\equiv\rho_0$, the rest distribution charge that generates the electromagnetic field\footnote{Proof: from $u^{\alpha}j_{\alpha}=u^0j_0-u^ij_i=\gamma^2c^2\rho_0-\gamma^2\rho_0\mathbf v\cdot\mathbf v\equiv\rho_0$ (where $\gamma=\left(1-v^2/c^2\right)^{-\frac12}$ is the relativistic factor).}, and $u^{\alpha}F_{\alpha\beta}=f_{\beta}$, the electromagnetic force per unit charge\footnote{$u^{\alpha}F_{\alpha0}$ is the electrical power given by the electric field to the test particle, and $u^{\alpha}F_{\alpha i}\equiv \gamma \left(\mathbf E +\mathbf v\times \mathbf B\right)$ is the Lorentz force per unit charge that acts on the particle.}:\begin{equation}\label{Efinal}
\mathcal{E}^{\left\{I\right\}}_{\mu\nu}=\dfrac {r\mu_0\rho_0}4h_{\mu\nu}+rf_{(\mu,\nu)}+\dfrac14r^2\mu_0T_{\mu\nu},
\end{equation}
which is the field equation of electromagnetism. In line with the last paragraph of Section \ref{I}, we see: in the second member of \ref{Efinal}, the first term represents the source of the electric charge; the second are derivatives of the electromagnetic force with respect to the coordinates; and the last is the energy-momentum density and the stress.

In particular, note that the first two terms at the right of \ref{Efinal} are contracted with the four velocity $u^{\alpha}$, and linearly scaled by the charge-to-mass ratio as $r^1$, while the last term is independent of the velocity and scales with $r^2$.


\subsection{Case II - $\mathcal F(x,u)=\alpha(x,u)+\beta(x,u)$}\label{CII}

This case encompasses equation \ref{Fh3} and is much more complex, presenting many higher-order terms. Evaluation of the spray coefficients and the curvatures is a little more difficult in this case, and we have left some key expressions to be presented in the Appendix. The final result is 
\begin{equation}\label{Gi-II}
G^{\mu}=\frac12 g^{\mu\sigma}\left(rF_{\lambda\sigma} u^{\lambda}+\frac{dU_{\sigma}}{ds}-F\partial_{\sigma}\alpha_0\right)
\end{equation}
for the spray coefficients. The first term on the second member of the above expression is equal to the result obtained for Case I (expression \ref{Gih}).

Replacing \ref{Gi-II} in \ref{Rmn} and \ref{Ric}, the results will be equal to \ref{Rmn3} and \ref{Rs} with additional intricated, higher-order terms, apparently without direct physical meaning, containing explicitly $\varphi_{\mu}$, and $U_{\mu}$  (see the Appendix, expression \ref{PhiU}), and derivatives. 


\section{Analysis of results}\label{Compar}

\subsection{Cases I and II}

Case I provided simple and physically meaningful expressions for $G_{\mu}$, $R_{\mu\nu}$, and $Ric$ and reveals that the suitable norm is \ref{Fh2}.

A particularly important finding is the emergence of the stress-energy tensor $T_{\mu\nu}$ in the field equation \ref{Efinal}.

\subsection{Comparison with the Riemannian geometry}

Finally, we may recall the Riemannian formalism of references \cite{D1,D2,D3} that culminated to the field equation \ref{Field}, where the charge-current tensor $Q_{\mu\nu}$ was separated in geometry $\Lambda_{\mu\nu}$, plus sources $\chi_{\mu\nu}$ \cite{D3}:
\begin{equation}
Q_{\mu\nu}=r^{-1}\Lambda_{\mu\nu}+\chi_{\mu\nu}.
\end{equation}
The comparison of terms shows that 
\begin{equation}
\left\{
\begin{array}{l}
\mathcal R=\chi_{\mu\nu}u^{\mu}u^{\nu}+\mathcal O(2),\\
\mathcal R_{\mu\nu}=\Lambda_{\mu\nu}+\mathcal O(2)
\end{array}
\right.
\end{equation}
in other words, the present formalism is identical to the formalism of RG up to first order.


\section*{Conclusions}\label{sec3}

The pseudo-Finsler geometry (Finsler geometry with non homogeneous norm) proved to be well-suited to the description of the electromagnetism, with the norm $\mathcal F=\sqrt{\eta_{\mu\nu}u^{\mu}u^{\nu}+2r\varphi_{\nu}u^{\nu}}$ (Case I).

The corresponding field equation \ref{Efinal} showed explicitly the energy-stress tensor $T_{\mu\nu}$, that is of course a feature shared with the Riemannian EFE of gravitation. 

The structure of the geodesic equation was preserved equal to the presented in references \cite{D1,D2,D3}.

Future studies may concern the definition of a timelike cone, the reality, smoothness and nondegeneracy that in the present low-field case were not considered.


\renewcommand{\theequation}{A.\arabic{equation}}

\setcounter{equation}{0}
\appendix
\section*{Appendix -- Mathematical support for the Case II}\label{Ap}

The present Appendix details the calculation of the derivatives necessary for the determination of the spray coefficients $G^{\mu}$, the Ricci curvature $R_{\mu\nu}$, and the scalar curvature $R$, as well as these magnitudes.

\subsection*{Auxiliary functions and notation:}

First of all,the well-known electromagnetic tensor:
\begin{equation}\label{Fmn}
F_{mu\nu}=\partial_{\mu}\varphi_{\nu}-\partial_{\nu}\varphi_{\mu}.
\end{equation}
Auxiliary functions:
\begin{equation}\label{PhiU}
\begin{array}{ll}
U_{\mu}=\eta_{\mu\nu}u^{\nu}, \ \ \ \ \ \ \alpha_0=\eta_{\mu\nu}u^{\mu}u^{\nu}\\
\end{array}
\end{equation}

Dot derivative (derivative with respect to a component $u^{\mu}$ of the four-velocity):
\begin{equation}
\dot{\partial}_{\mu}=\frac{\partial}{\partial u^{\mu}}
\end{equation}

\subsection*{Some derivatives:}

\begin{equation}\label{d-abF}
\left[
\begin{array}{l}
\partial_{\lambda} \mathcal F\approx \partial_{\lambda}\alpha_0+\partial_{\lambda}\beta\\
\dot{\partial}_{\sigma} F\approx U_{\sigma}+r\varphi_{\sigma}\\
\partial_{\lambda}\left(\dot{\partial}_{\sigma}\mathcal F^2\right)\approx 2\partial_{\lambda}\left(U_{\sigma}+r\varphi_{\sigma}\right).
\end{array}\\
\right.
\end{equation}
Then, the spray coefficient is:
\begin{equation}
G^{\mu}=\frac12 g^{\mu\sigma}\left(rF_{\lambda\sigma} u^{\lambda}+\left(\partial_{\lambda} U_{\sigma}\right)u^{\lambda}-F\partial_{\sigma}\alpha_0\right)
\end{equation}
However, $\left(\partial_{\lambda} U_{\sigma}\right)u^{\lambda}=dU_{\sigma}/ds$, then,
\begin{equation}
G^{\mu}=\frac12 g^{\mu\sigma}\left(rF_{\lambda\sigma} u^{\lambda}+\frac{dU_{\sigma}}{ds}-F\partial_{\sigma}\alpha_0\right)
\end{equation}

\end{document}